\documentclass[sigconf]{acmart}

\usepackage{multirow}
\usepackage{xspace}
\usepackage{amsmath}
\usepackage{appendix}
\usepackage{array}
\usepackage{tikz}
\usepackage{flushend}
\usepackage{balance}
\usepackage{microtype}
\usepackage{graphicx}
\usepackage{todonotes}
\usepackage{color}
\usepackage{listings}
\usepackage{multirow}
\usepackage{multicol}

\newcommand{\PreserveBackslash}[1]{\let\temp=\\#1\let\\=\temp}
\newcolumntype{C}[1]{>{\PreserveBackslash\centering}m{#1}}
\newcolumntype{R}[1]{>{\PreserveBackslash\raggedleft}m{#1}}
\newcolumntype{L}[1]{>{\PreserveBackslash\raggedright}m{#1}}
\newcommand*{\circled}[1]{\lower.7ex\hbox{\tikz\draw (0pt, 0pt)%
    circle (.5em) node {\makebox[1em][c]{\small #1}};}}

\AtBeginDocument{%
  \providecommand\BibTeX{{%
    \normalfont B\kern-0.5em{\scshape i\kern-0.25em b}\kern-0.8em\TeX}}}

\newcommand{\sys}{\textit{WhyGen}\xspace} % name of the system level benchmark
\newcommand{\ie}{\textit{i}.\textit{e}.,~}
\newcommand{\eg}{\textit{e}.\textit{g}.,~}

\begin{document}

\title{WhyGen: Explaining ML-powered Code Generation by Referring to Training Examples}

\author{Weixiang Yan}
\affiliation{%
  \institution{Beijing University of Posts and Telecommunications}
  \city{Beijing}
  \country{China}
}
\email{yanweixiang@bupt.edu.cn}

\author{Yuanchun Li}
\affiliation{%
  \institution{Institute for AI Industry Research (AIR), Tsinghua University}
  \city{Beijing}
  \country{China}}
\email{liyuanchun@air.tsinghua.edu.cn}

% \renewcommand{\shortauthors}{Weixiang Yan and Yuanchun Li.}

\begin{abstract}
Deep learning has demonstrated great abilities in various code generation tasks.
However, despite the great convenience for some developers, many are concerned that the code generators may recite or closely mimic copyrighted training data without user awareness, leading to legal and ethical concerns. To ease this problem, we introduce a tool, named \sys, to explain the generated code by referring to training examples. 
Specifically, we first introduce a data structure, named inference fingerprint, to represent the decision process of the model when generating a prediction.
The fingerprints of all training examples are collected offline and saved to a database. When the model is used at runtime for code generation, the most relevant training examples can be retrieved by querying the fingerprint database.
Our experiments have shown that \sys is able to precisely notify the users about possible recitations and highly similar imitations with a top-10 accuracy of 81.21$\%$.
The demo video can be found at \url{https://youtu.be/EtoQP6850To}.
\end{abstract}



% \czp{I cannot even find some of the keywords (e.g., database) in the abstract and introduction.}}

\maketitle

\section{Introduction}
\label{sec:intro}

Deep learning has recently been applied to various code generation tasks and has shown remarkable progress \cite{lu2021codexglue,wang2021codet5}.
For instance, GitHub Copilot \cite{chen2021codex}, a giant deep neural network developed by OpenAI, is able to generate highly-usable code from simple docstring or code prompts.
Such code generators can greatly improve the efficiency of developers by letting them focus on the high-level design rather than on the implementation details.

However, many developers are worried about the use of copyrighted source code for training such ML-powered code generators.
The machine learning models may have memorized the training data and generate code that is verbatim or very similar to the training examples.
Consequently, it may lead to licensing infringement if it generates and injects copyrighted code into customers' software.
% At the same time, the code generator occasionally recites the personal information in the training dataset exactly, which undoubtedly leads to privacy and security issues.

Although there has already been a lot of debates on this issue from the legal perspectives \cite{copilotCopyright,franceschelli2021copyright,pearce2021empirical}, how to technically ease this tension is still an open problem.
There is an inevitable trade-off between achieving higher accuracy and reducing training data memorization.
The success of today's DNN-powered code generators is largely due to their remarkable accuracy, and thus sacrificing the accuracy for less ethical concern may not be a sustainable solution.

We argue that a better way out is to keep the accurate training as it is, while additionally referring to the relevant training examples upon code generation.
On the one hand, the users of the code generators can understand why a certain code snippet is generated and learn more details from the referred examples (including the license and detailed usage).
On the other hand, the code generators do not need to sacrifice accuracy by reducing training data or memorization.
Achieving this goal is challenging since DNN models are usually regarded as black boxes that are very difficult to interpret.

% We believe that one of the major obstacles in code recitation research is the lack of measurement methods. Given a ML-powered code generator, it is easy to evaluate the model performance based on its test accuracy, but it is difficult to know whether and how the model has memorized the training data.

To this end, we introduce \sys, a tool to explain the predictions of ML-powered code generators by examples.
\sys solves the aforementioned problem by introducing a novel data structure, named inference fingerprint, to represent the design process of a model.
An inference fingerprint is a vector of activation values produced by a set of critical intermediate neurons in the network during the inference pass.
The fingerprint vectors can be compared across different inference passes, where similar samples would yield similar fingerprints.
Therefore, when the model is used online for code generation, we can compare the generated fingerprint with the fingerprints produced by the training examples, and retrieve the most relevant training examples to explain the generation.

We implement \sys on a popular open-source DNN-based code generator named CodeGPT \cite{lu2021codexglue} and test it on the PY150 dataset \cite{Raychev2016ProbabilisticMF}.
We randomly select 10,000 test examples that recite training data (\ie generating code snippets that are the same or very similar to uncommon training examples), and check whether \sys can find the recited examples at inference time.
We find that \sys can precisely locate the related training examples with a top-10 accuracy of 81.21$\%$. Meanwhile, the latency of retrieving related training examples is around 6 ms, which is minimal as compared to code generation.

The rest of this paper is organized as follows. Section~\ref{sec:method} introduces the design of the tool. Section~\ref{sec:evaluation} presents the accuracy and performance of the tool based on experiments. Section~\ref{sec:related} and Section~\ref{sec:conclusion} introduce the related work and future work, respectively.

\begin{figure*}
    \centering
    \includegraphics[width=5.4in]{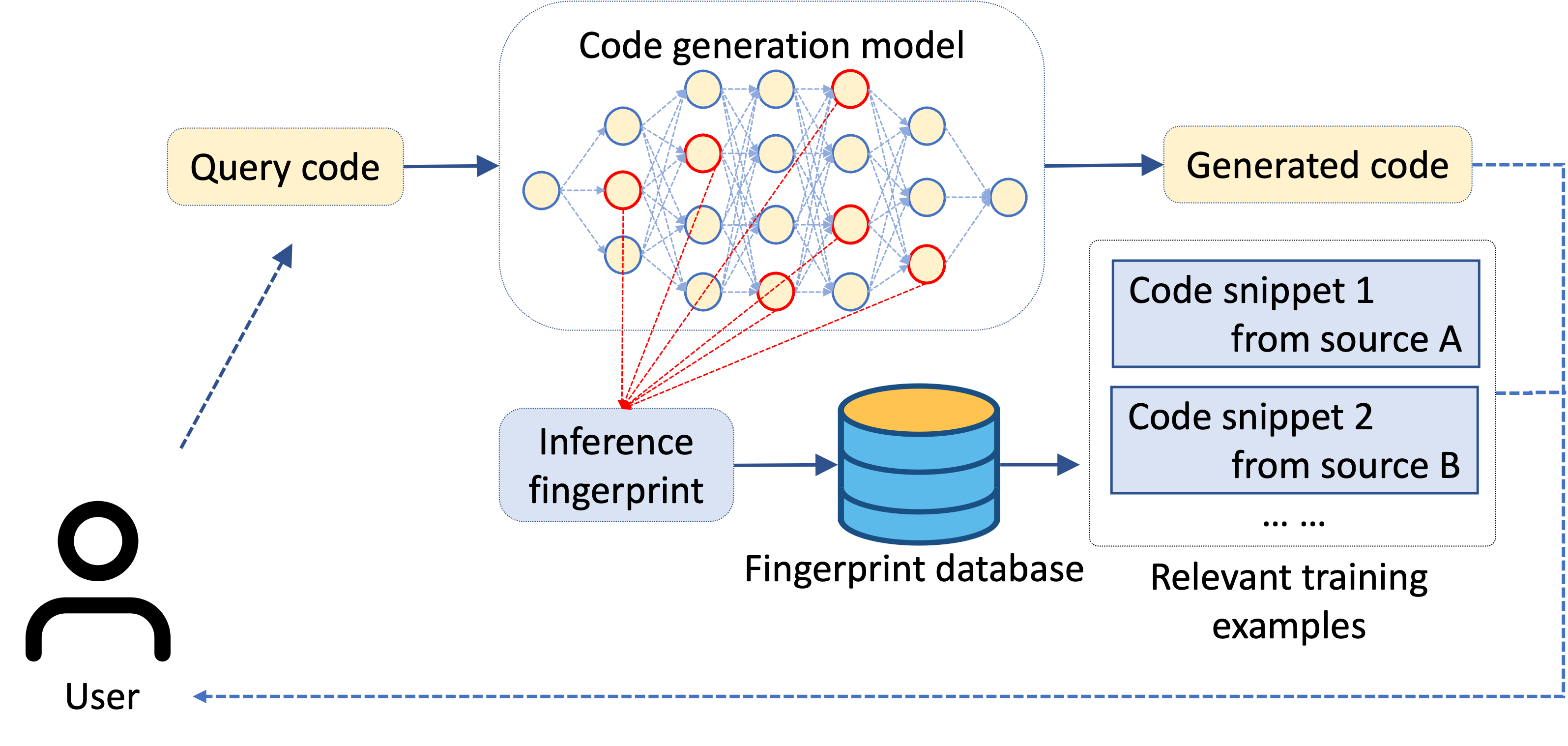}
    \caption{The workflow of \sys to explain DNN-powered code generation by examples.}
    \label{fig:workflow}
\end{figure*}

\section{Tool Design}
\label{sec:method}

The workflow of \sys is shown in Figure~\ref{fig:workflow}. For each query code given by the user (a programmer who is using the ML-powered code generator), we extract an inference fingerprint from the neural network. The fingerprint is used to query a fingerprint dataset to find the most similar fingerprints and their corresponding training examples. The retrieved training examples are then returned to the user with the code generated by the model, giving them prompts about which training examples are potentially relevant to the current generation. We also provide the source (\eg the link to the original GitHub repository) of each relevant training example to the user for further reference.

\subsection{Inference Fingerprint}
\label{sec:fingerprint_gen}

Understanding which training samples are more relevant to a certain generation is challenging, because neural networks are usually regarded as black boxes that are difficult to interpret.
The training examples are used to compute gradients that accumulate into millions of model weights.
It is hard to distinguish the contribution of each training example after the model parameters are learned.

Instead of analyzing which training examples contribute the most to the code generation, we analyze which training examples trigger similar decision logic as the user query.
We assume the training examples with similar decision logic are the relevant examples for the generated code.
This assumption, though not formally provable, is intuitive because human brains also process relevant concepts with similar decision pattern.

We introduce a data structure, named \emph{inference fingerprint}, to represent the decision logic of the neural network and compare across different data examples.
An inference fingerprint is a vector of activation values produced by a set of intermediate neurons in the network during the inference pass.
The same set of intermediate neurons is used to produce the fingerprints, and thus the fingerprints are comparable across different data examples.
Prior work has attempted to use intermediate neurons to represent the decision logic of DNN \cite{zhang2020dynamic,liu2020pmc}, but they are mainly designed for other purposes (such as adversarial detection, data distribution estimation, etc.) and the computation of critical neurons is relatively slow.

In our work, the selection of the intermediate neurons for producing fingerprints must meet two objectives.
First, the number of selected intermediate neurons must be small, since the total number of neurons in a code generator model is too huge to compute.
%The excessive number of neurons will increase the time consumption of \sys, thus damaging the user experience.
Second, the selected intermediate neurons should be representative, so that the relevant code examples can be grouped together.

\begin{figure}
    \centering
    \includegraphics[width=2.8in]{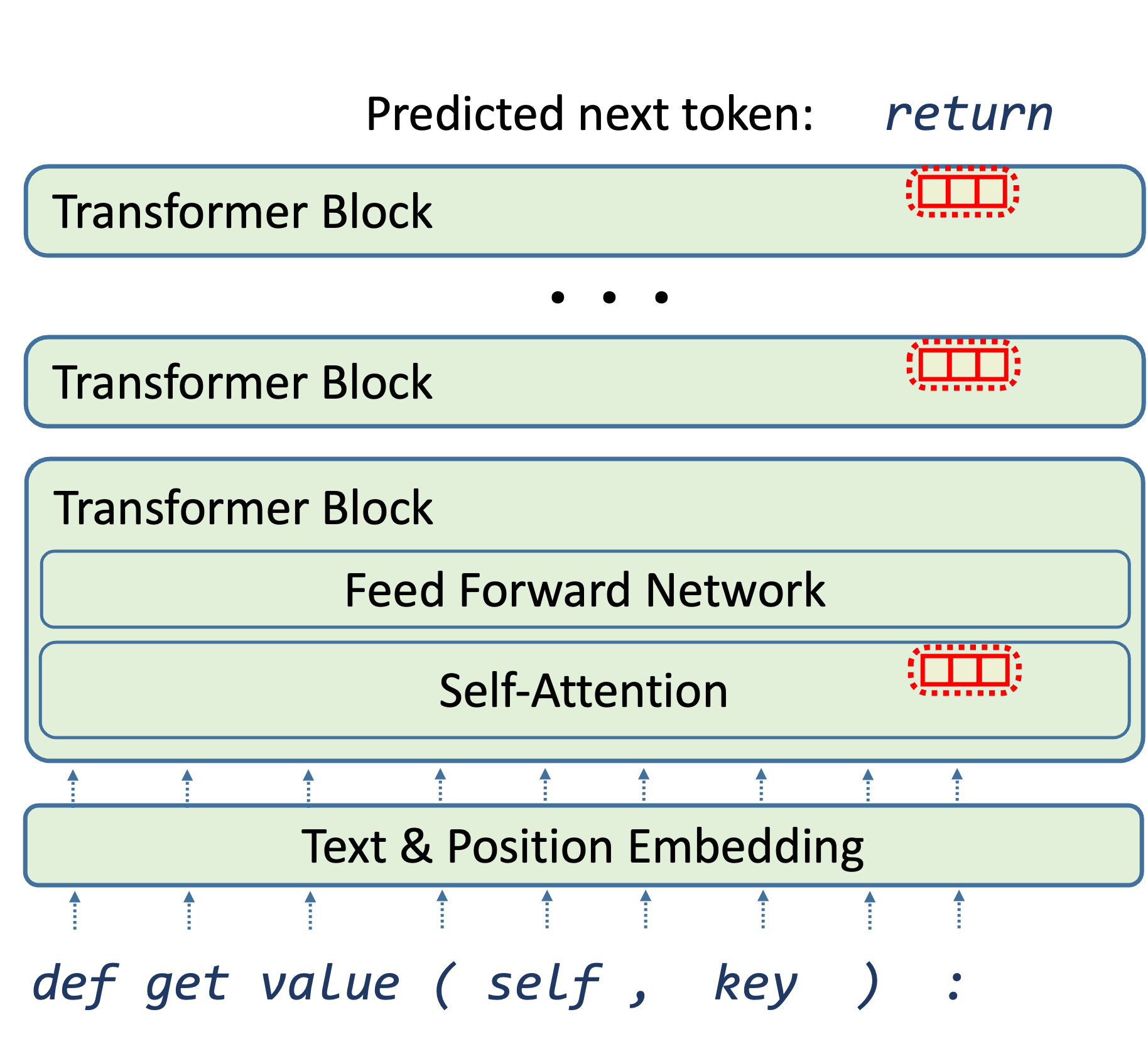}
    \caption{Illustration of an inference step with Transformer-based code generator. The intermediate neurons to compute fingerprints are selected from the activation layers corresponding to the first generated token.}
    \label{fig:transformer_neurons}
\end{figure}

Modern code generators are mostly based on the Transformer architecture \cite{lu2021codexglue,wang2021codet5,ahmad2021unified}.A typical inference step of a Transformer-based code generator is illustrated in Figure~\ref{fig:transformer_neurons}, in which the input is a sequence of preceding code tokens, and the output is the predicted next token.
Each piece of generated code is produced in a token-by-token manner, where each token is predicted by an inference step. 
The predicted token in a step is appended to the query sequence and used as the input to predict the subsequent token in the next step.

Taking CodeGPT \cite{lu2021codexglue} as an example, it takes a sequence of tokens as the input and predicts the next token step by step until the \texttt{<end>} identifier is predicted. In each step of next-token prediction, CodeGPT uses the Beam Search algorithm to retain the top-k candidate tokens with the highest scores. Then for each of these top-k candidates, it further runs the inference pass and finds the top-k highest-score candidate tokens, resulting in $K^{2}$ candidate combinations. Among them, only the top-k candidate combinations with the highest scores are kept in the next step, and the process repeats until the end of decoding. Finally, the candidate token combination with the highest score is returned as the final prediction.

We combine the heuristic understanding of the model and quantitative methods to locate the intermediate neurons.
We first narrow down the selection of intermediate neurons to the activation layers after each encoder module, because they are designed as the result of each independent encoding stage.
Moreover, we focus on the activation values corresponding to the first generated token since they have encoded all user-generated input tokens and are more explicitly related to the generated code.

To further locate the neurons that can better represent the decision process, we use a profiling phase to understand the behavior of the neurons in activation layers.
The training samples are fed into the model and the neuron output values are recorded. We compute several statistics based on the profiling results and compare several criteria to select the critical neurons. We find that the most high-variance neurons are more representative, and their output values are concatenated together as the inference fingerprint.

\subsection{Training Data Indexing and Retrieval}

Next, we compute the inference fingerprints for all training examples and save them to a database.
% \textcolor{blue}{ 
The inference fingerprint generation process for the training examples is consistent with the process for user input (as described in Section~\ref{sec:fingerprint_gen}),
in order to ensure that the inference fingerprints of training examples can be compared and searched with the fingerprint generated by the user input at the test time.
% In Section 3.2, in different baseline settings, they are all consistent with each other.}
Each record in the database includes the inference fingerprint, the code snippet, and the original source (\eg repository URL and/or file path) of the code.
The fingerprint vectors are indexed to speed up the process of searching for the most relevant training examples.

When the code generator produces a prediction, we compute the inference fingerprint for the prediction, and find the most similar fingerprints in the database. The similarity is measured as the Euclidean distance between the two vectors.
The training examples corresponding to the most similar inference fingerprints are returned to the user as the relevant training examples.

\subsection{Implementation Details}

We implement the prototype of \sys with an open-source DNN-powered code generator CodeGPT \cite{lu2021codexglue}, which is based on an advanced language model GPT-2 \cite{radford2019language} and fine-tuned on the PY150 dataset \cite{Raychev2016ProbabilisticMF}.
The state-of-the-art closed-source code generator, Codex or Copilot \cite{chen2021codex}, is based on GPT-3 architecture. While larger in size, GPT-3 is conceptually and structurally similar to GPT-2. Thus, we believe our method can be applied to it as well.

To index and search for the fingerprints, we use the Faiss open-source library \cite{faiss}.
The size of the inference fingerprint is set to 100 in our implementation, and the number of returned relevant training examples is set to 10 by default.

% \begin{figure*}[ht] 
% \centering 
% \includegraphics[scale=0.11]{figures/BSC search.jpg} 
% \caption{The most similar code search process based on a search engine, which is divided into three steps: establishing a search engine, BSC query request, and returning and calculating BSC.} 
% \label{fig1} 
% \end{figure*}

\section{Evaluation}
\label{sec:evaluation}

% \ycl{First, explain how we compute the recitation rate CodeGPT and CodeT5.}

% \ycl{Then, do some breakdown analysis, \eg the distribution of the lengths of recited code, the categories of the recited code, etc. Show some figures and/or tables here.}

% \ycl{Finally, apply some simple methods, such as output randomization and differential privacy, and see the influence on the recitation rate and model accuracy.}

We conduct experiments to evaluate \sys in terms of effectiveness (whether it can generate meaningful relevant training examples) and overhead (how much time it needs to retrieve the relevant examples).

\subsection{Experiment Setup}

Since the relevance of training examples is a subjective concept, directly evaluating it is difficult. Thus, we take an indirect approach instead - we first find some reciting behaviors of the code generator (\ie the generator generates code exactly the same as in the training set). The recitations are regarded as the ground truth of relevant examples, so the effectiveness of \sys can be evaluated by examining whether the recited code snippets appear in the results produced by \sys.

To find the recitations, we randomly pick 10,000 code snippets from the test set and use the code generator to predict the next line for each snippet. 
For each predicted line of code, we search the training dataset to find the most similar line, \ie the line with the shortest edit distance to the predicted line. If the edit distance is 0 and the code line is unique enough (number of occurrences is smaller than 10), we consider it as a recitation. In the end, we obtain 3,842 cases of recitations.
% Then we evaluate \sys by examine whether the relevant examples generated by \sys can successfully include the recited training examples.
We use the top-k accuracy metric to evaluate \sys, which means the probability that the recited training example is among the top k examples returned by \sys.

\subsection{Effectiveness of \sys}

\begin{table}[]
    \centering
    \begin{tabular}{c|c|c|c}
    \toprule
        Method & Acc@10 & Acc@5 & Acc@1 \\
    \midrule
        \sys &  \textbf{81.21$\%$} &  \textbf{79.28$\%$} &  \textbf{73.84$\%$}\\ 
        Random & 67.57$\%$ & 66.61$\%$ & 62.78$\%$\\ 
        Maximum & 56.32$\%$ & 54.89$\%$ & 51.09$\%$\\ 
        Minimum & 57.26$\%$ & 55.62$\%$ & 52.32$\%$\\ 
        FFN & 79.46$\%$ & 77.98$\%$ & 73.43$\%$\\
    \bottomrule
    \end{tabular}
    \caption{The accuracy of \sys and its variants to include the recited code in the retrieved training examples.}
    \label{tab:accuracy}
\end{table}

%The accuracy@k result of \sys is shown in Table~\ref{tab:accuracy}. \ycl{explain the results}

Based on the found recitations, we evaluate the effectiveness of \sys.
Due to the lack of baselines in this area, we compare the default configuration of \sys with several variants. Each variant uses a different strategy to select the critical neurons to compute the inference fingerprint. For example, ``Random'' means to randomly select the intermediate neurons, ``Maximum'' and ``Minimum'' mean to select the neurons with maximum or minimum output values, and ``FFN'' means to select high-variance neurons from the feed-forward network layer rather than the self-attention layer.

The accuracy results are shown in Table~\ref{tab:accuracy}.
% \textcolor{blue}{Table 1 shows the results of reciting or highly imitating training examples in the training examples returned by several methods.
% We not only provide the default accuracy of top-10, but also provide more stringent accuracy of top-5 and top-1. 
% In addition to the \sys method we proposed, we also added Random, Maximum, Minimum, and FFN as references. 
Clearly, our default configuration of \sys achieves the best results with a top-10 accuracy of 81.21\% and top-1 accuracy of 73.84\%, which is significantly better than using other criteria to select the fingerprint neurons.
Selecting critical neurons from the FFN layer can achieve competitive results, but it is still slightly less effective than using the self-attention layers.

The accuracy results imply that the inference fingerprint computed by \sys does a good job in encoding important information about the decision-making process during the code generation, and it can effectively be used to find the training samples that share the similar decision logic with the query sample.

% indicates that \sys is not only good at recording important information in the decision-making process of generating predictions, but also better at finding the training samples that have the greatest weight of influence on a given prediction result.

% \textcolor{blue}{Based on the above results, we use \sys to conduct a case study during the inference process when entering a long enough context on CodeGPT.
% Figure~\ref{fig:Whygen_case_study} shows the results returned by \sys on a test example, where Query Code is the input of CodeGPT and \sys, and then CodeGPT gives the prediction result of the next line code.
% At the same time, \sys returned the most responsible top-5 relevant training examples in the training samples and the file path of their source.}

\begin{figure}
    \centering
    \includegraphics[width=3.4in]{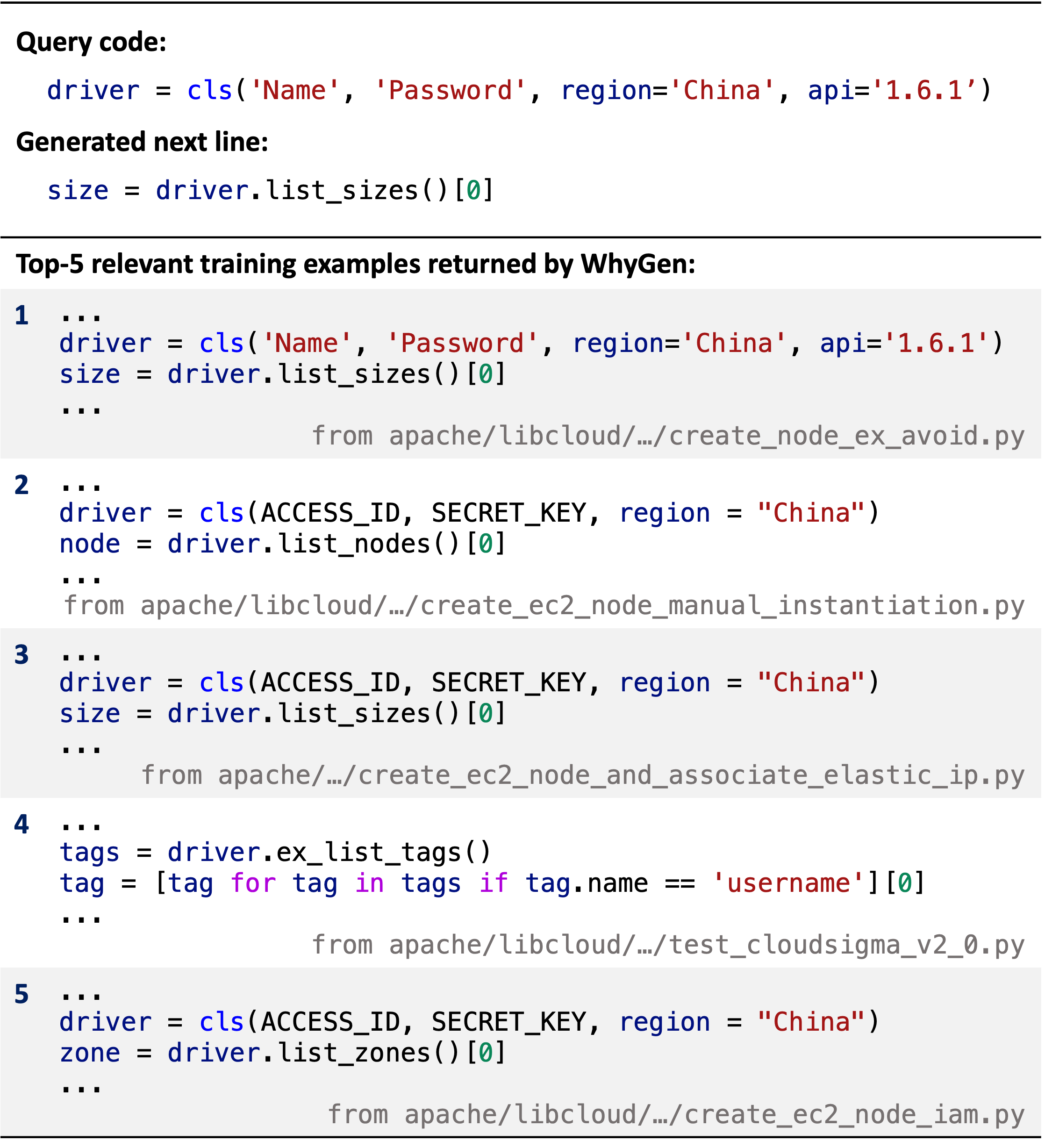}
    \caption{An example of relevant training examples returned by \sys.}
    \vspace{-0.1cm}
    \label{fig:Whygen_case_study}
\end{figure}

Figure~\ref{fig:Whygen_case_study} shows an example of relevant training examples returned by \sys when generating the next line for a given query code.
We can see that the returned five training examples are almost all very relevant to the query code and generated code, and the first example is an exact recitation.
% \textcolor{blue}{We found that if the prediction given by the code generator is recited or highly imitated, \sys can accurately match it to the relevant training examples. 
% In other words, \sys can track the recited or imitated training samples in the returned examples when recited or highly imitated behaviors occur in the code generator. 
In practice, the returned relevant training examples can serve as a reminder or guidance for the user. If the generated code recites or highly imitates the copyrighted code, the user can modify or abandon the generated code to avoid legal and ethical concerns.
\sys will also provide the source path of the returned training examples, so that users can learn more about the code predicted by the code generator and decide whether to use it in their own software.

%Table~\ref{tab:accuracy} illustrate some examples of relevant training examples generated by \sys. \ycl{explain, describe interesting findings}

\subsection{Overhead of \sys}

We further measure the overhead of \sys in training and serving scenarios using a Linux server with an AMD EPYC 7742 CPU.

In the training stage, \sys needs to compute the fingerprints for all training examples and build an index for the fingerprints. The whole process takes around 20 hours, which is shorter than the training time of code generator models (around 25 hours). We believe the training overhead is acceptable since it is a one-time offline cost.

In the serving stage, \sys needs to compute the inference fingerprint and retrieve relevant examples for each prediction made by the code generator. The overhead is around 6 ms, which is minimal as compared to the code generation process (360 ms). Thus, we believe our tool can be used in real-time to give meaningful prompts to the code generator users.

\section{Related Work}
\label{sec:related}

\textbf{Instance-based Model Interpretation.} %Introduce the ICML work (and its related work) with 2-3 sentences. Describe their disadvantages (overhead).
Interpreting deep neural networks with training examples has become one of the major methods for model interpretation.
The most representative instance-based interpretation technique is the influence function approach \cite{koh2017understanding}, which traces a model’s predictions through its learning algorithm and back to the training data using influence functions.
% It mainly uses the inference training input and the change of the weight of the training samples, then the prediction of the model is tracked through the learning algorithm, and finally determines the training sample that has the greatest impact on the given prediction. 
However, the calculation of the influence function is very computationally intensive, making it difficult even impossible to be applied to large language models and datasets.
% The influence function needs to calculate the Hessian matrix involving all parameters, but for large neural networks with 10 billion parameters such as CodeT5 \cite{wang2021codet5}, it is too expensive and difficult to calculate.

\textbf{Privacy leakage in language models.} 
The training example recitation problem in code generators is similar to the privacy leakage problem in language models, which has been discussed intensively in prior work \cite{pan2020privacyRisksLM,inan2021leakageAnalysis,carlini2020extracting}.
In order to reduce such privacy concerns, a common solution is using differential privacy techniques \cite{abadi2016dldp}, \ie adding noise during training to avoid memorizing individual details.
However, applying differential privacy may significantly harm model accuracy, specifically for large language models \cite{bagdasaryan2019differential}.

% Taking GitHub copilot as an example, it not only encounters the problem of code infringement, but also faces the problem of privacy disclosure because the training dataset contains personal data. Copilot claims on its official website that its internal test results show that there are few personal data completely consistent with the training set in the prediction code segment given by copilot. Sometimes, copilot will still recite personal data such as e-mail address and telephone number in the prediction code segment, but these data are synthesized based on the patterns in the training data. In the technical preview, copilot implements a filter to eliminate the phenomenon of personal data Recitation in the prediction code segment. However, the facts are not what copilot claims. When a software engineer used copilot to generate the prediction code segment, the personal information of another developer was recited in the prediction code segment. Such privacy and security issues are also what we want to avoid. WX: This paragraph seems to be more appropriate in the background motivation section.

% \textbf{Code copyright infringement.}
% The intellectual property issues of source code have been concerned by the community for a long time.
% For example, a lot of research efforts have been put on code plagiarism detection. \xx
% Copyright infringement of machines has been rarely discussed before, and it now becomes an urgent problem due to the rapid development of deep learning technology.
% This work aim to advance this thread of research .

\section{Conclusion and Future Work}
\label{sec:conclusion}

We introduce a tool to explain the code generated by DNN models by referring to training examples.
The tool can possibly be used as an IDE plugin along with the auto-completion feature. We hope our technique can help reduce the concern about using unauthorized source code for training code generators.

As future work, we plan to improve the accuracy of retrieving relevant training examples by exploring better inference fingerprints. 
We also plan to extend \sys to support more and larger code generators based on the Transformer architecture and other architectures such as CNN and RNN, in order to ensure good generalizability and practicability of \sys.
A larger and more standard benchmark would be useful to better evaluate different training examples retrieving methods.
Moreover, it would be interesting and helpful to investigate better quantitative metrics to measure the causal relationship between the training examples and the generated code, which can be used to evaluate \sys and other explain-by-example techniques more comprehensively and rigorously.

%It would also be meaningful to understand the causality between the training examples and the generated code.
Our tool is open-sourced at \url{https://github.com/WeixiangYAN/WhyGen}.

% \section*{Acknowledgment}

\bibliographystyle{ACM-Reference-Format}
\balance
\bibliography{ref}


\begin{thebibliography}{18}

%%% the Web address.

\ifx \showCODEN    \undefined \def \showCODEN     #1{\unskip}     \fi
\ifx \showDOI      \undefined \def \showDOI       #1{#1}\fi
\ifx \showISBNx    \undefined \def \showISBNx     #1{\unskip}     \fi
\ifx \showISBNxiii \undefined \def \showISBNxiii  #1{\unskip}     \fi
\ifx \showISSN     \undefined \def \showISSN      #1{\unskip}     \fi
\ifx \showLCCN     \undefined \def \showLCCN      #1{\unskip}     \fi
\ifx \shownote     \undefined \def \shownote      #1{#1}          \fi
\ifx \showarticletitle \undefined \def \showarticletitle #1{#1}   \fi
\ifx \showURL      \undefined \def \showURL       {\relax}        \fi
% invisible to TeX
\providecommand\bibfield[2]{#2}
\providecommand\bibinfo[2]{#2}
\providecommand\natexlab[1]{#1}
\providecommand\showeprint[2][]{arXiv:#2}

\bibitem[\protect\citeauthoryear{??}{cop}{2021}]%
        {copilotCopyright}
 \bibinfo{year}{2021}\natexlab{}.
\newblock \bibinfo{title}{GitHub, Copilot and the Copyright Around AI}.
\newblock
  \bibinfo{howpublished}{https://www.plagiarismtoday.com/2021/07/08/github-copilot-and-the-copyright-around-ai/}.
\newblock


\bibitem[\protect\citeauthoryear{Abadi, Chu, Goodfellow, McMahan, Mironov,
  Talwar, and Zhang}{Abadi et~al\mbox{.}}{2016}]%
        {abadi2016dldp}
\bibfield{author}{\bibinfo{person}{Martin Abadi}, \bibinfo{person}{Andy Chu},
  \bibinfo{person}{Ian Goodfellow}, \bibinfo{person}{H~Brendan McMahan},
  \bibinfo{person}{Ilya Mironov}, \bibinfo{person}{Kunal Talwar}, {and}
  \bibinfo{person}{Li Zhang}.} \bibinfo{year}{2016}\natexlab{}.
\newblock \showarticletitle{Deep learning with differential privacy}. In
  \bibinfo{booktitle}{\emph{Proceedings of the 2016 ACM SIGSAC conference on
  computer and communications security}}. \bibinfo{pages}{308--318}.
\newblock


\bibitem[\protect\citeauthoryear{Ahmad, Chakraborty, Ray, and Chang}{Ahmad
  et~al\mbox{.}}{2021}]%
        {ahmad2021unified}
\bibfield{author}{\bibinfo{person}{Wasi~Uddin Ahmad}, \bibinfo{person}{Saikat
  Chakraborty}, \bibinfo{person}{Baishakhi Ray}, {and} \bibinfo{person}{Kai-Wei
  Chang}.} \bibinfo{year}{2021}\natexlab{}.
\newblock \bibinfo{title}{Unified Pre-training for Program Understanding and
  Generation}.
\newblock
\newblock
\showeprint[arxiv]{2103.06333}~[cs.CL]


\bibitem[\protect\citeauthoryear{Bagdasaryan, Poursaeed, and
  Shmatikov}{Bagdasaryan et~al\mbox{.}}{2019}]%
        {bagdasaryan2019differential}
\bibfield{author}{\bibinfo{person}{Eugene Bagdasaryan}, \bibinfo{person}{Omid
  Poursaeed}, {and} \bibinfo{person}{Vitaly Shmatikov}.}
  \bibinfo{year}{2019}\natexlab{}.
\newblock \showarticletitle{Differential privacy has disparate impact on model
  accuracy}.
\newblock \bibinfo{journal}{\emph{Advances in Neural Information Processing
  Systems}}  \bibinfo{volume}{32} (\bibinfo{year}{2019}),
  \bibinfo{pages}{15479--15488}.
\newblock


\bibitem[\protect\citeauthoryear{Carlini, Tramer, Wallace, Jagielski,
  Herbert-Voss, Lee, Roberts, Brown, Song, Erlingsson, et~al\mbox{.}}{Carlini
  et~al\mbox{.}}{2020}]%
        {carlini2020extracting}
\bibfield{author}{\bibinfo{person}{Nicholas Carlini}, \bibinfo{person}{Florian
  Tramer}, \bibinfo{person}{Eric Wallace}, \bibinfo{person}{Matthew Jagielski},
  \bibinfo{person}{Ariel Herbert-Voss}, \bibinfo{person}{Katherine Lee},
  \bibinfo{person}{Adam Roberts}, \bibinfo{person}{Tom Brown},
  \bibinfo{person}{Dawn Song}, \bibinfo{person}{Ulfar Erlingsson},
  {et~al\mbox{.}}} \bibinfo{year}{2020}\natexlab{}.
\newblock \showarticletitle{Extracting training data from large language
  models}.
\newblock \bibinfo{journal}{\emph{arXiv preprint arXiv:2012.07805}}
  (\bibinfo{year}{2020}).
\newblock


\bibitem[\protect\citeauthoryear{Chen, Tworek, Jun, Yuan, Ponde, Kaplan,
  Edwards, Burda, Joseph, Brockman, et~al\mbox{.}}{Chen et~al\mbox{.}}{2021}]%
        {chen2021codex}
\bibfield{author}{\bibinfo{person}{Mark Chen}, \bibinfo{person}{Jerry Tworek},
  \bibinfo{person}{Heewoo Jun}, \bibinfo{person}{Qiming Yuan},
  \bibinfo{person}{Henrique Ponde}, \bibinfo{person}{Jared Kaplan},
  \bibinfo{person}{Harri Edwards}, \bibinfo{person}{Yura Burda},
  \bibinfo{person}{Nicholas Joseph}, \bibinfo{person}{Greg Brockman},
  {et~al\mbox{.}}} \bibinfo{year}{2021}\natexlab{}.
\newblock \showarticletitle{Evaluating large language models trained on code}.
\newblock \bibinfo{journal}{\emph{arXiv preprint arXiv:2107.03374}}
  (\bibinfo{year}{2021}).
\newblock


\bibitem[\protect\citeauthoryear{Franceschelli and Musolesi}{Franceschelli and
  Musolesi}{2021}]%
        {franceschelli2021copyright}
\bibfield{author}{\bibinfo{person}{Giorgio Franceschelli} {and}
  \bibinfo{person}{Mirco Musolesi}.} \bibinfo{year}{2021}\natexlab{}.
\newblock \bibinfo{title}{Copyright in Generative Deep Learning}.
\newblock
\newblock
\showeprint[arxiv]{2105.09266}~[cs.CY]


\bibitem[\protect\citeauthoryear{Inan, Ramadan, Wutschitz, Jones, R{\"u}hle,
  Withers, and Sim}{Inan et~al\mbox{.}}{2021}]%
        {inan2021leakageAnalysis}
\bibfield{author}{\bibinfo{person}{Huseyin~A Inan}, \bibinfo{person}{Osman
  Ramadan}, \bibinfo{person}{Lukas Wutschitz}, \bibinfo{person}{Daniel Jones},
  \bibinfo{person}{Victor R{\"u}hle}, \bibinfo{person}{James Withers}, {and}
  \bibinfo{person}{Robert Sim}.} \bibinfo{year}{2021}\natexlab{}.
\newblock \showarticletitle{Training data leakage analysis in language models}.
\newblock \bibinfo{journal}{\emph{arXiv preprint arXiv:2101.05405}}
  (\bibinfo{year}{2021}).
\newblock


\bibitem[\protect\citeauthoryear{Johnson, Douze, and J{\'e}gou}{Johnson
  et~al\mbox{.}}{2017}]%
        {faiss}
\bibfield{author}{\bibinfo{person}{Jeff Johnson}, \bibinfo{person}{Matthijs
  Douze}, {and} \bibinfo{person}{Herv{\'e} J{\'e}gou}.}
  \bibinfo{year}{2017}\natexlab{}.
\newblock \showarticletitle{Billion-scale similarity search with GPUs}.
\newblock \bibinfo{journal}{\emph{arXiv preprint arXiv:1702.08734}}
  (\bibinfo{year}{2017}).
\newblock


\bibitem[\protect\citeauthoryear{Koh and Liang}{Koh and Liang}{2017}]%
        {koh2017understanding}
\bibfield{author}{\bibinfo{person}{Pang~Wei Koh} {and} \bibinfo{person}{Percy
  Liang}.} \bibinfo{year}{2017}\natexlab{}.
\newblock \showarticletitle{Understanding black-box predictions via influence
  functions}. In \bibinfo{booktitle}{\emph{International Conference on Machine
  Learning}}. PMLR, \bibinfo{pages}{1885--1894}.
\newblock


\bibitem[\protect\citeauthoryear{Liu, Li, Liu, Guo, and Chen}{Liu
  et~al\mbox{.}}{2020}]%
        {liu2020pmc}
\bibfield{author}{\bibinfo{person}{Bingyan Liu}, \bibinfo{person}{Yuanchun Li},
  \bibinfo{person}{Yunxin Liu}, \bibinfo{person}{Yao Guo}, {and}
  \bibinfo{person}{Xiangqun Chen}.} \bibinfo{year}{2020}\natexlab{}.
\newblock \showarticletitle{Pmc: A privacy-preserving deep learning model
  customization framework for edge computing}.
\newblock \bibinfo{journal}{\emph{Proceedings of the ACM on Interactive,
  Mobile, Wearable and Ubiquitous Technologies}} \bibinfo{volume}{4},
  \bibinfo{number}{4} (\bibinfo{year}{2020}), \bibinfo{pages}{1--25}.
\newblock


\bibitem[\protect\citeauthoryear{Lu, Guo, Ren, Huang, Svyatkovskiy, Blanco,
  Clement, Drain, Jiang, Tang, et~al\mbox{.}}{Lu et~al\mbox{.}}{2021}]%
        {lu2021codexglue}
\bibfield{author}{\bibinfo{person}{Shuai Lu}, \bibinfo{person}{Daya Guo},
  \bibinfo{person}{Shuo Ren}, \bibinfo{person}{Junjie Huang},
  \bibinfo{person}{Alexey Svyatkovskiy}, \bibinfo{person}{Ambrosio Blanco},
  \bibinfo{person}{Colin Clement}, \bibinfo{person}{Dawn Drain},
  \bibinfo{person}{Daxin Jiang}, \bibinfo{person}{Duyu Tang}, {et~al\mbox{.}}}
  \bibinfo{year}{2021}\natexlab{}.
\newblock \showarticletitle{CodeXGLUE: A Machine Learning Benchmark Dataset for
  Code Understanding and Generation}.
\newblock \bibinfo{journal}{\emph{arXiv preprint arXiv:2102.04664}}
  (\bibinfo{year}{2021}).
\newblock


\bibitem[\protect\citeauthoryear{Pan, Zhang, Ji, and Yang}{Pan
  et~al\mbox{.}}{2020}]%
        {pan2020privacyRisksLM}
\bibfield{author}{\bibinfo{person}{Xudong Pan}, \bibinfo{person}{Mi Zhang},
  \bibinfo{person}{Shouling Ji}, {and} \bibinfo{person}{Min Yang}.}
  \bibinfo{year}{2020}\natexlab{}.
\newblock \showarticletitle{Privacy risks of general-purpose language models}.
  In \bibinfo{booktitle}{\emph{2020 IEEE Symposium on Security and Privacy
  (SP)}}. IEEE, \bibinfo{pages}{1314--1331}.
\newblock


\bibitem[\protect\citeauthoryear{Pearce, Ahmad, Tan, Dolan-Gavitt, and
  Karri}{Pearce et~al\mbox{.}}{2021}]%
        {pearce2021empirical}
\bibfield{author}{\bibinfo{person}{Hammond Pearce}, \bibinfo{person}{Baleegh
  Ahmad}, \bibinfo{person}{Benjamin Tan}, \bibinfo{person}{Brendan
  Dolan-Gavitt}, {and} \bibinfo{person}{Ramesh Karri}.}
  \bibinfo{year}{2021}\natexlab{}.
\newblock \bibinfo{title}{An Empirical Cybersecurity Evaluation of GitHub
  Copilot's Code Contributions}.
\newblock
\newblock
\showeprint[arxiv]{2108.09293}~[cs.CR]


\bibitem[\protect\citeauthoryear{Radford, Wu, Child, Luan, Amodei, Sutskever,
  et~al\mbox{.}}{Radford et~al\mbox{.}}{[n.\,d.]}]%
        {radford2019language}
\bibfield{author}{\bibinfo{person}{Alec Radford}, \bibinfo{person}{Jeffrey Wu},
  \bibinfo{person}{Rewon Child}, \bibinfo{person}{David Luan},
  \bibinfo{person}{Dario Amodei}, \bibinfo{person}{Ilya Sutskever},
  {et~al\mbox{.}}} \bibinfo{year}{[n.\,d.]}\natexlab{}.
\newblock \showarticletitle{Language models are unsupervised multitask
  learners}.
\newblock  (\bibinfo{year}{[n.\,d.]}).
\newblock


\bibitem[\protect\citeauthoryear{Raychev, Bielik, and Vechev}{Raychev
  et~al\mbox{.}}{2016}]%
        {Raychev2016ProbabilisticMF}
\bibfield{author}{\bibinfo{person}{Veselin Raychev}, \bibinfo{person}{Pavol
  Bielik}, {and} \bibinfo{person}{Martin~T. Vechev}.}
  \bibinfo{year}{2016}\natexlab{}.
\newblock \showarticletitle{Probabilistic model for code with decision trees}.
\newblock \bibinfo{journal}{\emph{Proceedings of the 2016 ACM SIGPLAN
  International Conference on Object-Oriented Programming, Systems, Languages,
  and Applications}} (\bibinfo{year}{2016}).
\newblock


\bibitem[\protect\citeauthoryear{Wang, Wang, Joty, and Hoi}{Wang
  et~al\mbox{.}}{2021}]%
        {wang2021codet5}
\bibfield{author}{\bibinfo{person}{Yue Wang}, \bibinfo{person}{Weishi Wang},
  \bibinfo{person}{Shafiq Joty}, {and} \bibinfo{person}{Steven~CH Hoi}.}
  \bibinfo{year}{2021}\natexlab{}.
\newblock \showarticletitle{CodeT5: Identifier-aware Unified Pre-trained
  Encoder-Decoder Models for Code Understanding and Generation}.
\newblock \bibinfo{journal}{\emph{arXiv preprint arXiv:2109.00859}}
  (\bibinfo{year}{2021}).
\newblock


\bibitem[\protect\citeauthoryear{Zhang, Li, Guo, Chen, and Liu}{Zhang
  et~al\mbox{.}}{2020}]%
        {zhang2020dynamic}
\bibfield{author}{\bibinfo{person}{Ziqi Zhang}, \bibinfo{person}{Yuanchun Li},
  \bibinfo{person}{Yao Guo}, \bibinfo{person}{Xiangqun Chen}, {and}
  \bibinfo{person}{Yunxin Liu}.} \bibinfo{year}{2020}\natexlab{}.
\newblock \showarticletitle{Dynamic slicing for deep neural networks}. In
  \bibinfo{booktitle}{\emph{Proceedings of the 28th ACM Joint Meeting on
  European Software Engineering Conference and Symposium on the Foundations of
  Software Engineering}}. \bibinfo{pages}{838--850}.
\newblock


\end{thebibliography}

\end{document}